\documentclass[a4paper]{jpconf}

\usepackage{graphicx}
\usepackage{amssymb}
\begin{document}
\title{The M{\sc ajorana} experiment:  an ultra-low background search for neutrinoless double-beta decay}

\author{D~G~Phillips~II$^{1,2}$, E~Aguayo$^{3}$, F~T~Avignone~III$^{4,5}$, H~O~Back$^{6,2}$, A~S~Barabash$^{7}$, M~Bergevin$^{8}$, F~E~Bertrand$^{5}$, M~Boswell$^{9}$, V~Brudanin$^{10}$, M~Busch$^{11,2}$, Y-D~Chan$^{8}$, C~D~Christofferson$^{12}$, J~I~Collar$^{13}$, D~C~Combs$^{6,2}$, R~J~Cooper$^{5}$, J~A~Detwiler$^{8}$, P~J~Doe$^{15}$, Y~Efremenko$^{14}$, V~Egorov$^{10}$, H~Ejiri$^{16}$, S~R~Elliott$^{9}$, J~Esterline$^{11,2}$, J~E~Fast$^{3}$, N~Fields$^{13}$, P~Finnerty$^{1,2}$, F~M~Fraenkle$^{1,2}$, V~M~Gehman$^{9}$, G~K~Giovanetti$^{1,2}$, M~P~Green$^{1,2}$, V~E~Guiseppe$^{17}$, K~Gusey$^{10}$, A~L~Hallin$^{18}$, R~Hazama$^{16}$, R~Henning$^{1,2}$, A~Hime$^{9}$, E~W~Hoppe$^{3}$, M~Horton$^{12}$, S~Howard$^{12}$, M~A~Howe$^{1,2}$, R~A~Johnson$^{15}$, K~J~Keeter$^{19}$, C~Keller$^{17}$, M~F~Kidd$^{9}$, A~Knecht$^{15}$, O~Kochetov$^{10}$, S~I~Konovalov$^{7}$, R~T~Kouzes$^{3}$, B~LaFerriere$^{3}$, B~H~LaRoque$^{9}$, J~Leon$^{15}$, L~E~Leviner$^{6,2}$, J~C~Loach$^{8}$, S~MacMullin$^{1,2}$, M~G~Marino$^{15}$, R~D~Martin$^{8}$, D~-M~Mei$^{17}$, J~Merriman$^{3}$, M~L~Miller$^{15}$, L~Mizouni$^{4,3}$, M~Nomachi$^{16}$, J~L~Orrell$^{3}$, N~R~Overman$^{3}$, A~W~P~Poon$^{8}$, G~Perumpilly$^{17}$, G~Prior$^{8}$, D~C~Radford$^{5}$, K~Rielage$^{9}$, R~G~H~Robertson$^{15}$, M~C~Ronquest$^{9}$, A~G~Schubert$^{15}$, T~Shima$^{16}$, M~Shirchenko$^{10}$, K~J~Snavely$^{1,2}$, D~Steele$^{9}$, J~Strain$^{1,2}$, K~Thomas$^{17}$, V~Timkin$^{10}$, W~Tornow$^{11,2}$, I~Vanyushin$^{7}$, R~L~Varner$^{5}$, K~Vetter$^{20,8}$, K~Vorren$^{1,2}$, J~F~Wilkerson$^{1,2,5}$, B~A~Wolfe$^{15}$, E~Yakushev$^{10}$, A~R~Young$^{6,2}$, C~-H~Yu$^{5}$, V~Yumatov$^{7}$ and C~Zhang$^{17}$}

\address{$^{1}$~Department of Physics and Astronomy, University of North Carolina, Chapel Hill, NC, USA}
\address{$^{2}$~Triangle Universities Nuclear Laboratory, Durham, NC, USA}
\address{$^{3}$~Pacific Northwest National Laboratory, Richland, WA, USA}
\address{$^{4}$~Department of Physics and Astronomy, University of South Carolina, Columbia, SC, USA}
\address{$^{5}$~Oak Ridge National Laboratory, Oak Ridge, TN, USA}
\address{$^{6}$~Department of Physics, North Carolina State University, Raleigh, NC, USA}
\address{$^{7}$~Institute for Theoretical and Experimental Physics, Moscow, Russia}
\address{$^{8}$~Nuclear Science Division, Lawrence Berkeley National Laboratory, Berkeley, CA, USA}
\address{$^{9}$~Los Alamos National Laboratory, Los Alamos, NM, USA}
\address{$^{10}$~Joint Institute for Nuclear Research, Dubna, Russia}
\address{$^{11}$~Department of Physics, Duke University, Durham, NC, USA}
\address{$^{12}$~South Dakota School of Mines and Technology, Rapid City, SD, USA}
\address{$^{13}$~Department of Physics, University of Chicago, Chicago, IL, USA}
\address{$^{14}$~Department of Physics and Astronomy, University of Tennessee, Knoxville, TN, USA}
\address{$^{15}$~Center for Experimental Nuclear Physics and Particle Astrophysics, and Department of Physics, University of Washington, Seattle, WA, USA}
\address{$^{16}$~Research Center for Nuclear Physics, and Department of Physics, Osaka University, Ibaraki, Osaka, Japan}
\address{$^{17}$~Department of Physics, University of South Dakota, Vermillion, SD, USA}
\address{$^{18}$~Center for Particle Physics, University of Alberta, Edmonton, AB, Canada}
\address{$^{19}$~Department of Physics, Black Hills State University, Spearfish, SD, USA}
\address{$^{20}$~Alternate address:  Department of Nuclear Engineering, University of California, Berkeley, CA, USA}

\ead{dgp@email.unc.edu}

\begin{abstract}
The observation of neutrinoless double-beta decay would resolve the Majorana nature of the neutrino and could provide information on the absolute scale of the neutrino mass.  The initial phase of the M{\sc ajorana} experiment, known as the D{\sc emonstrator}, will house 40 kg of Ge in an ultra-low background shielded environment at the 4850' level of the Sanford Underground Laboratory in Lead, SD.  The objective of the D{\sc emonstrator} is to determine whether a future 1-tonne experiment can achieve a background goal of one count per tonne-year in a narrow region of interest around the $^{76}$Ge neutrinoless double-beta decay peak.  
\end{abstract}

\section{Introduction}

The primary goal of the M{\sc ajorana} experiment is to determine the Majorana or Dirac nature of the neutrino by performing an ultra-low background search for neutrinoless double-beta decay (0$\nu\beta\beta$) using high-purity Ge (HPGe) as a source and a detector.  Based on observations of neutrino oscillations in a number of experiments employing a wide variety of different techniques, we now know that neutrinos have non-zero masses.  Two independent squared mass differences ($\Delta m_{12}^{2}$, $\Delta m_{23}^{2}$) composed of the three-neutrino family mass parameters and three mixing angles ($\theta_{13}$, $\theta_{12}$, $\theta_{23}$) provide an adequate explanation of established neutrino oscillation results [1].  While neutrino oscillation experiments are sensitive to neutrino mass splittings, they are generally not useful probes of absolute neutrino masses.  The identity of the neutrino as a Majorana particle and valuable information on the absolute scale of the neutrino mass would be revealed from observation of 0$\nu\beta\beta$ [2, 3, 4, 5, 6].  0$\nu\beta\beta$ would also be a definitive indication of lepton number violation [7].  The low threshold of the detectors employed by the M{\sc ajorana} experiment will also allow for a search for low-mass weakly interacting massive particle (WIMP) dark matter.      

\section{The M{\sc ajorana} Demonstrator}

The initial phase of the M{\sc ajorana} experiment, known as the D{\sc emonstrator}, will house 40 kg of Ge in an ultra-low background shielded environment composed of ultra-pure electroformed copper and lead.  The D{\sc emonstrator} will be located at the Davis campus on the 4850' level of the Sanford Underground Laboratory in Lead, SD, in an area that includes a cleanroom detector laboratory, a cleanroom electroforming laboratory and a cleanroom machine shop.  Up to 30 kg of the Ge employed for the D{\sc emonstrator} will be enriched to 86$\%$ in $^{76}$Ge.  The objective of the D{\sc emonstrator} is to achieve a background rate of four counts per tonne-year in a 4 keV region of interest (ROI)  around the $^{76}$Ge 0$\nu\beta\beta$ Q-value at 2039 keV.   In a future tonne-scale experiment, better granularity combined with a deeper location and thicker electroformed copper shield would scale the aforementioned background rate to one count per tonne-year.  For a tonne-scale experiment, a background rate of less than one count per tonne-year in the ROI would yield a half-life sensitivity of $T^{0\nu}_{1/2} > 10^{27}$ years (see \hbox{Fig.} 1) and allow for the exploration of the inverted mass hierarchy [8].  After a year of running, the D{\sc emonstrator} should be able to set a limit on the half-life sensitivity of $T^{0\nu}_{1/2} > 4\times 10^{25}$ years and address the claim of 0$\nu\beta\beta$ detection in $^{76}$Ge [9].   

\begin{figure}[h]
  \begin{center}
    \scalebox{0.7}{\includegraphics{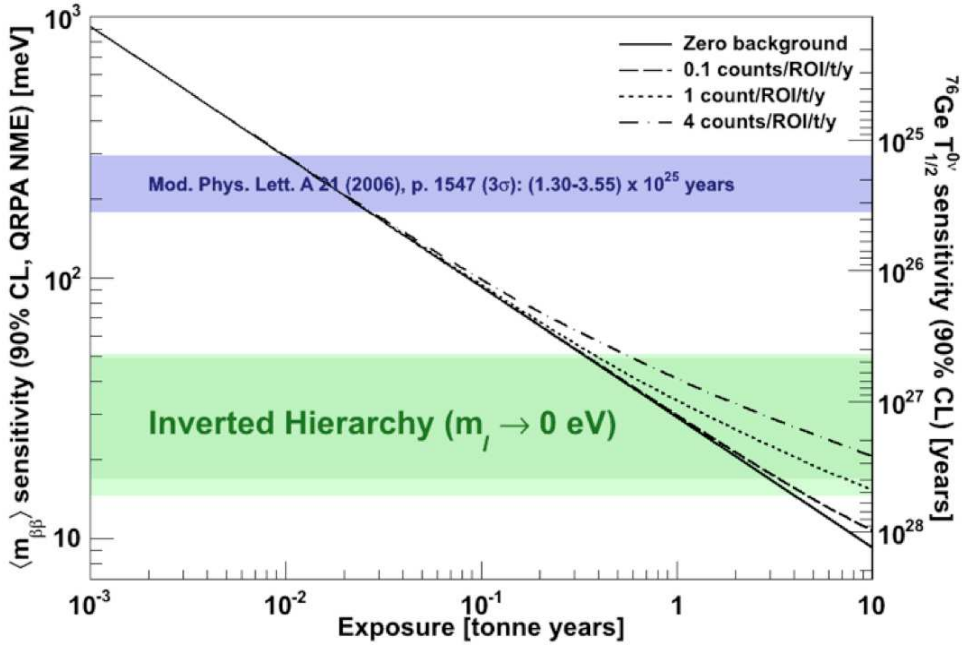}}
    \caption{Sensitivity of searching for 0$\nu\beta\beta$ using Ge enriched to 86$\%$ in $^{76}$Ge as a function of exposure and background [10].  The set of values required for refuting or confirming a recent observational claim of 0$\nu\beta\beta$ [9] and addressing inverted hierarchy are shown.  The dark green area for inverted hierarchy encompasses the full range of CP-violating phases, while the light green area includes uncertainties in measured neutrino oscillation parameters.}
  \end{center}
\end{figure}

P-type Point Contact (PPC) detectors [11, 12] have been chosen as the detector technology for the M{\sc ajorana} D{\sc emonstrator}.  PPC detectors have an extremely small signal electrode compared to standard semi-coaxial HPGe detectors.  They also feature a low detector capacitance on the order of 1 pF and low leakage current, thereby producing a detector with very low electronic noise and an energy threshold less than 1 keV.  The superlative performance of PPC detectors in the low energy regime will enable the M{\sc ajorana} D{\sc emonstrator} to perform a search for low-mass, WIMP dark matter [12, 13, 14] in addition to the primary experimental objective of probing for 0$\nu\beta\beta$.  The low thresholds of the M{\sc ajorana} PPC detectors will allow for detection of recoiling WIMPs with masses of $\sim$10 GeV [15].  $^{238}$U, $^{232}$Th, and their decay chain daughters can create backgrounds in the 0$\nu\beta\beta$ ROI, such as Compton-scattering of gamma rays within the Ge detectors.  The two electrons in $^{76}$Ge $\beta\beta$-decays typically deposit ionization energy within a 1-2 mm range in the Ge detector, whereas Compton-scattered gamma rays deposit energy over cm scales.  The PPC detector geometry is ideally suited to the task of isolating the 0$\nu\beta\beta$ events from Compton-scattered backgrounds.  PPC detectors feature longer drift times than standard HPGe semi-coaxial detectors and a rapid rise in weighing potential near the point contact, which allows for excellent discrimination between 0$\nu\beta\beta$-like single-site events and multi-site background events using pulse-shape analysis (PSA) techniques [16].  A typical HPGe detector resolution at the $^{76}$Ge 0$\nu\beta\beta$ Q-value of 0.2$\%$ is more than adequate for distinguishing between 2$\nu\beta\beta$ and 0$\nu\beta\beta$ events in $^{76}$Ge.

The M{\sc ajorana} D{\sc emonstrator} will deploy the 40 kg of Ge detectors in two ultra-pure electroformed Cu cryostats (see \hbox{Fig.} 2 and \hbox{Fig.} 3).  In the summer of 2012, a prototype cryostat of two strings of natural Ge detectors will be deployed in a surface laboratory at Los Alamos National Laboratory (LANL).  The two electroformed Cu cryostats will house seven strings of five PPC detectors apiece, which will be mounted in electroformed Cu components.  Deployment of the first cryostat of natural and enriched Ge detectors will occur underground at Sanford Underground Laboratory (SUL) by the summer of 2013, while the second cryostat is expected to be deployed underground at SUL in 2014.    Backgrounds in the ROI are expected from cosmogenic impurities, such as $^{68}$Ge in the PPC detectors and $^{60}$Co in the PPC detectors and copper shielding.  Electroforming copper and minimizing surface exposure improves the radiopurity of cryostat and mounting components to a level that will satisfy the stringent background requirements of the M{\sc ajorana} experiment.  M{\sc ajorana} has already begun electroforming operations with ten baths in a cleanroom laboratory at the 4850' level Ross shop area of SUL and six baths in a shallow underground cleanroom laboratory at Pacific Northwest National Laboratory in Richland, WA, where a Th activity level in electroformed copper of $<$0.7 $\mu$Bq/kg has been demonstrated [17].  Radon influx in the baths is mitigated with nitrogen cover gas.  The electroforming bath solution is composed of high-purity Optima grade acids and high-purity water ($>$18 M$\Omega$/cm$^{3}$) with growth rates of $\sim$0.07 mm/day.     

Enrichment of Ge is taking place at the electrochemical plant in Zelonogorsk, Russia.  An initial delivery of $\sim$ 30 kg of enriched GeO$_{2}$ was shipped from Russia using ground-transportation in a steel-shielded container and has arrived in Oak Ridge, TN where it is being stored in an underground location with 120 \hbox{ft.} of rock overburden.   The GeO$_{2}$ will undergo processing into electronic-grade material before further refinement into detector grade material using zone-refining and Czochralski crystal-growing techniques.  After the PPC detectors are manufactured, they will be moved underground into a cleanroom lab at the 4850' level Davis campus of SUL.  At the Davis campus lab, the detectors will undergo characterization testing before they are mounted in an electroformed copper cryostat and moved into the shield.  The cryostats will be surrounded by an electroformed copper track for movement of sources during calibration runs.  A 5 cm inner layer of electroformed copper, a 5 cm middle layer of oxygen-free high-conductivity copper and a 45 cm outer layer of commercial lead surrounded by 30 cm of polyethylene will make up the shield volume.  The copper and lead shield will be hermetically sealed and injected with nitrogen to mitigate radon influx.  A plastic scintillator cosmic-ray veto will surround the shield and be located inside the polyethylene.    

\begin{figure}[h]
\begin{minipage}{17pc}
\includegraphics[width=13pc]{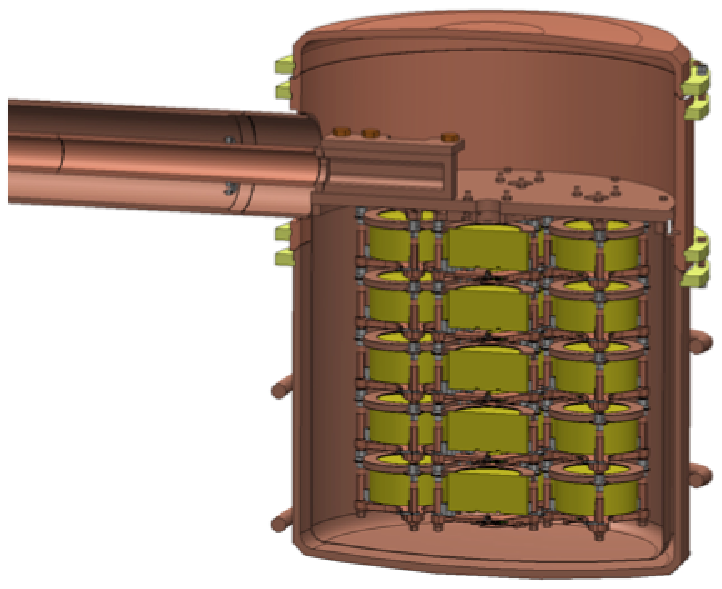}
\caption{\label{label}Cross-sectional view of a M{\sc ajorana} electroformed copper cryostat.}
\end{minipage}\hspace{2.5pc}%
\begin{minipage}{15pc}
\includegraphics[width=19pc]{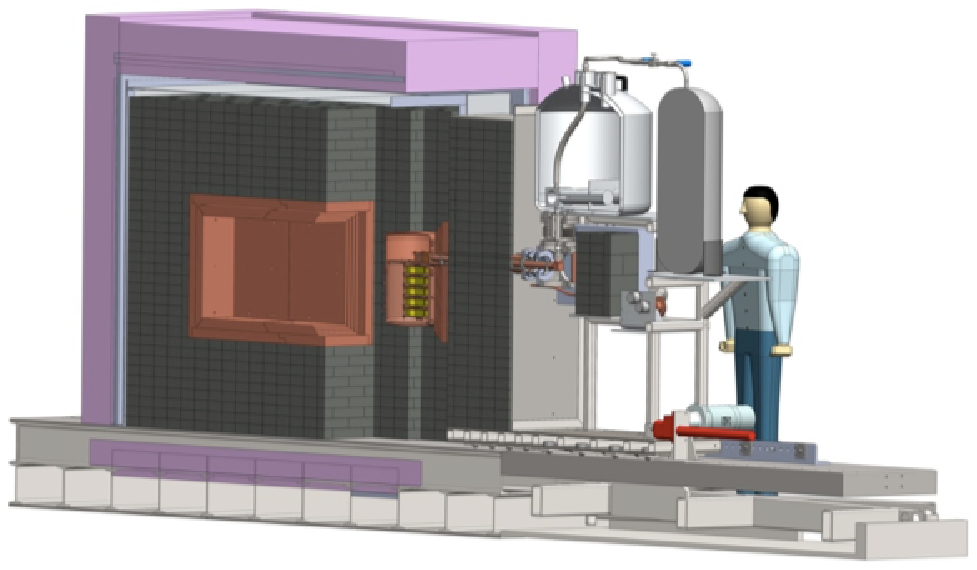}
\caption{\label{label}Cross-sectional view of the M{\sc ajorana} D{\sc emonstrator}.}
\end{minipage} 
\end{figure}

\section{Detector Validation and Characterization}

The 40 kg of Ge used for the D{\sc emonstrator} will be modified versions of commercially available Canberra PPC-like Broad Energy Germanium (BEGe) detectors, where up to 30 kg will be enriched to 86$\%$ in $^{76}$Ge.   A number of  M{\sc ajorana} D{\sc emonstrator} natural Ge modified BEGe detectors and PPC detectors have already undergone a rigorous validation and characterization campaign [12, 16, 18].  All modified BEGe detectors that have undergone testing have exceeded specifications for leakage current of 30 pA, detector capacitance of 2.0 pF and nominal relative detection efficiency of 34$\%$.  Thirty-five of the modified BEGe detectors have achieved FWHM below specifications for an energy resolution of 2.2 keV at the 1332.5 keV emission line of $^{60}$Co.  Events in which a $\gamma$-ray pair from positron annihilation escape the detector without depositing any energy are referred to as double-escape peak events.  Since the resulting energy deposition from the two charged leptons is highly localized they exhibit the single-site behavior referred to earlier.  Since multi-site events typically involve one photon that deposits its entire energy in the detector, these events are also referred to as full-energy peak events.  The pulse-shape performance for a subset of the tested PPC-type detectors have been studied with the survival probability of full-energy peak events \hbox{vs.} double-escape peak events in these detectors given in \hbox{Fig.} 4.  A modified BEGe detector with a passivation ditch radius of 15 mm and a 4 mm point contact is being studied at the Kimballton Underground Research Facility (KURF) at a depth of 1450 meters water equivalent.  The capacitance of the modified BEGe at KURF has been measured to be $\sim$1.55 pF and the detector threshold has been demonstrated to be $\sim$1.0 keV [19].  All PPC-type detectors that have been tested meet the pulse-shape discrimination requirements for the M{\sc ajorana} D{\sc emonstrator} project.  Table 1 lists the evaluation status of all M{\sc ajorana} PPC and modified BEGe detectors.  

\begin{figure}[h]
  \begin{center}
    \scalebox{0.65}{\includegraphics{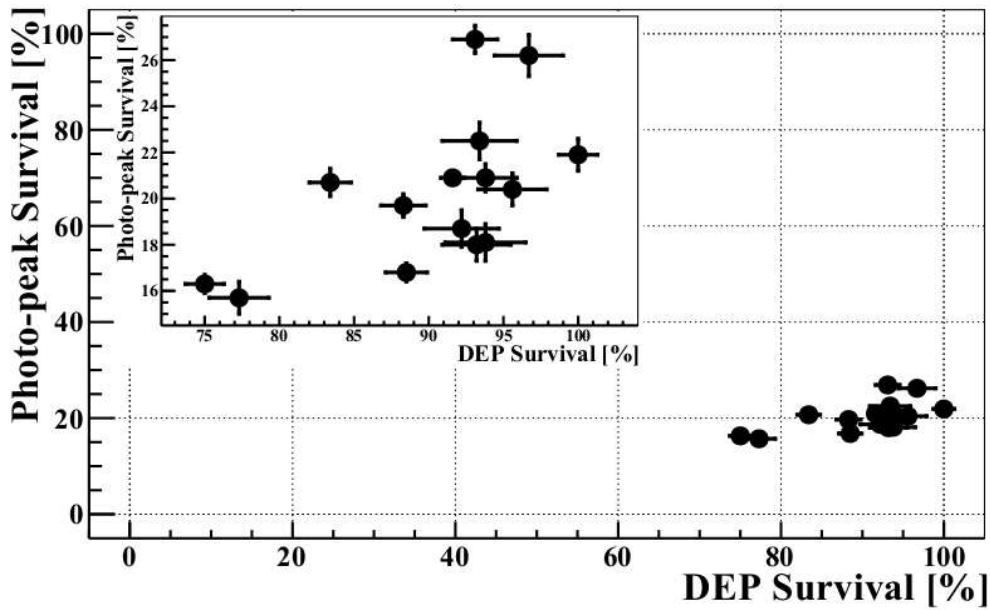}}
    \caption{Survival probability of full-energy peak events (multi-site) vs. 0$\nu\beta\beta$ signal-like double-escape peak events (single-site) for a set of PPC-type detectors [18].}
  \end{center}
\end{figure}

\begin{table} \caption{\label{label}M{\sc ajorana} D{\sc emonstrator} PPC and modified BEGe detector evaluations. Note that `u.e.' stands for `under evaluation'.} \begin{center}
\begin{tabular}{llllll}
\br
Manufacturer&Type&Quantity&Diameter (mm)&Length (mm)&Reference\\
\mr
Canberra USA&modified BEGe&43&70&30&[18], u.e.\\
Canberra USA&modified BEGe&2&60&30&[19], u.e.\\
Canberra USA&modified BEGe&1&90&30&[16]\\
Canberra France&modified BEGe&1&50&44&[12], u.e.\\
Canberra France&PPC&1&50&50&u.e.\\
ORTEC&PPC&3&67&54&[18], u.e.\\
ORTEC&PPC&1&62&51&[18]\\
PGT&PPC&1&70&30&[18]\\
PHDs&PPC&1&72&37&[18]\\
PHDs&PPC&1&62&46&u.e.\\
LBNL&PPC&1&62&50&u.e.\\
LBNL&PPC&3&20&10&u.e.\\
LBNL&segmented PPC&1&62&50&u.e.\\
LBNL&NPC&1&58&58&[11]\\
\br
\end{tabular}
\end{center}
\end{table}

\section{Simulations}

An extensive simulation campaign has been undertaken by the M{\sc ajorana} collaboration to estimate purity requirements for materials and verify analysis cuts.   All major components of the D{\sc emonstrator} geometry have been simulated.  The full U and Th chains have been simulated for all 3782 components of the experimental setup.  Simulations of $^{40}$K and $^{60}$Co have been performed for all metal components of the D{\sc emonstrator}.  The signal response of PPC detectors has been estimated by simulations of pulse-generation and charge drift inside the detectors [16].  

\section{Summary}

The M{\sc ajorana} D{\sc emonstrator} will search for 0$\nu\beta\beta$ using 40 kg of Ge in an ultra-low background environment at the 4850' level of SUL in Lead, SD.  A cryostat of natural and enriched Ge detectors will be deployed underground at the SUL Davis campus by the summer of 2013, while 2014 is expected to mark deployment of a second cryostat of Ge detectors underground at the SUL Davis campus.  Two strings of natural Ge detectors will be deployed in a prototype cryostat at LANL in the summer of 2012.  The M{\sc ajorana} D{\sc emonstrator} should be able to verify or refute the recent observational claim of 0$\nu\beta\beta$ [9] within 2-3 years of commissioning the first underground cryostat.


\section*{References}
\begin{thebibliography}{9}
\bibitem{1} Camilleri~L, Lisi~E and Wilkerson~J~F 2008 {\it Ann. Rev. of Nucl. and Part. Phys.} {\bf 58} 343-369
\bibitem{2} Schechter~J and Valle~J~W~F 1982 {\it Phys. Rev. D} {\bf 25} 2951-2954
\bibitem{3} Elliott~S~R and Vogel~P 2002 {\it Ann. Rev. of Nucl. and Part. Phys.} {\bf 52} 115-151
\bibitem{4} Barabash~A~S 2004 {\it Phys. Atom. Nucl.} {\bf 67} 438-452
\bibitem{5} Avignone~F~T~III, King~G~S and Zdesenko~Y~G 2005 {\it New J. Phys.} {\bf 7} 6
\bibitem{6} Avignone~F~T~III, Elliott~S~R and Engel~J 2008 {\it Rev. Mod. Phys.} {\bf 80} 481-516
\bibitem{7} Elliott~S~R and Engel~J 2004 {\it J. Phys. G} {\bf 30} R183
\bibitem{8} Rodin~V~A, Faessler~A, \v{S}imkovic~F and Vogel~P 2006 {\it Nucl. Phys. Lett. A} {\bf 766} 107-131 erratum 2007 {\it Nucl. Phys. A} {\bf 793} 213-215
\bibitem{9} Klapdor-Kleingrothaus~H~V and Krivosheina~I~V 2006 {\it Mod. Phys. Lett. A} {\bf 21} 1547-1566
\bibitem{10} \v{S}imkovic~F, Faessler~A, M\"{u}ther~H, Rodin V and Stauf~M 2009 {\it Phys. Rev. C} {\bf 79} 055501
\bibitem{11} Luke~P~N, Goulding~F~S, Madden~N~W and Pehl~R~H 1989 {\it IEEE Trans. on Nucl. Sci.} {\bf 36} 926
\bibitem{12} Barbeau~P~S, Collar~J~I and Tench~O 2007 {\it J. Cosm. and Astropart. Phys.} {\bf 09} 009
\bibitem{13} Aalseth~C~E {\it et al.} 2008 {\it Phys. Rev. Lett.} {\bf 101} 251301; 2009 {\it Phys. Rev. Lett.} {\bf 102} 109903(E)
\bibitem{14} Aalseth~C~E {\it et al.} 2011 {\it Phys. Rev. Lett.} {\bf 106} 131301
\bibitem{15} Marino M G 2010 Dark matter physics with p-type point-contact germanium detectors: extending the physics reach of the M{\sc ajorana} experiment {\it University of Washington Ph. D. Dissertation}
\bibitem{16} Cooper~R~J, Radford~D~C, Lagergren~K, Colaresi~J~F, Darken~L, Henning~R, Marino~M~G and Yocum~K~M 2011 {\it Nucl. Inst. and Meth. in Phys. Res. A} {\bf 629} 303-310
\bibitem{17} Hoppe~E~W {\it et al.} 2008 {\it J. of Radioanalytical and Nucl. Chem.} {\bf 277} 103-110
\bibitem{18} Gehman~V {\it et al.} 2011 P-type Point Contact Detectors for the M{\sc ajorana} D{\sc emonstrator} {\it Manuscript in Preparation}
\bibitem{19} Aalseth~C~E {\it et al.} 2011 {\it Nucl. Inst. and Meth. in Phys. Res. A} {\bf 652} 692-695

\end{thebibliography}
\end{document}